\documentclass[%
 reprint,
 amsmath,amssymb,
 aps,
prl,
superscriptaddress,
]{revtex4-2}

\usepackage{graphicx}
\usepackage{dcolumn}
\usepackage{bm}
\usepackage{booktabs}
\usepackage{gensymb}

\usepackage{hyperref}
\hypersetup{colorlinks=true,allcolors=blue}

\usepackage[all]{hypcap}

\usepackage[caption=false]{subfig}

\usepackage{lineno}

\usepackage{color}
\begin{document}

\title{First Search for Unstable Sterile Neutrinos with the IceCube Neutrino Observatory}

\affiliation{III. Physikalisches Institut, RWTH Aachen University, D-52056 Aachen, Germany}
\affiliation{Department of Physics, University of Adelaide, Adelaide, 5005, Australia}
\affiliation{Dept. of Physics and Astronomy, University of Alaska Anchorage, 3211 Providence Dr., Anchorage, AK 99508, USA}
\affiliation{Dept. of Physics, University of Texas at Arlington, 502 Yates St., Science Hall Rm 108, Box 19059, Arlington, TX 76019, USA}
\affiliation{CTSPS, Clark-Atlanta University, Atlanta, GA 30314, USA}
\affiliation{School of Physics and Center for Relativistic Astrophysics, Georgia Institute of Technology, Atlanta, GA 30332, USA}
\affiliation{Dept. of Physics, Southern University, Baton Rouge, LA 70813, USA}
\affiliation{Dept. of Physics, University of California, Berkeley, CA 94720, USA}
\affiliation{Lawrence Berkeley National Laboratory, Berkeley, CA 94720, USA}
\affiliation{Institut f{\"u}r Physik, Humboldt-Universit{\"a}t zu Berlin, D-12489 Berlin, Germany}
\affiliation{Fakult{\"a}t f{\"u}r Physik {\&} Astronomie, Ruhr-Universit{\"a}t Bochum, D-44780 Bochum, Germany}
\affiliation{Universit{\'e} Libre de Bruxelles, Science Faculty CP230, B-1050 Brussels, Belgium}
\affiliation{Vrije Universiteit Brussel (VUB), Dienst ELEM, B-1050 Brussels, Belgium}
\affiliation{Department of Physics and Laboratory for Particle Physics and Cosmology, Harvard University, Cambridge, MA 02138, USA}
\affiliation{Dept. of Physics, Massachusetts Institute of Technology, Cambridge, MA 02139, USA}
\affiliation{Dept. of Physics and The International Center for Hadron Astrophysics, Chiba University, Chiba 263-8522, Japan}
\affiliation{Department of Physics, Loyola University Chicago, Chicago, IL 60660, USA}
\affiliation{Dept. of Physics and Astronomy, University of Canterbury, Private Bag 4800, Christchurch, New Zealand}
\affiliation{Dept. of Physics, University of Maryland, College Park, MD 20742, USA}
\affiliation{Dept. of Astronomy, Ohio State University, Columbus, OH 43210, USA}
\affiliation{Dept. of Physics and Center for Cosmology and Astro-Particle Physics, Ohio State University, Columbus, OH 43210, USA}
\affiliation{Niels Bohr Institute, University of Copenhagen, DK-2100 Copenhagen, Denmark}
\affiliation{Dept. of Physics, TU Dortmund University, D-44221 Dortmund, Germany}
\affiliation{Dept. of Physics and Astronomy, Michigan State University, East Lansing, MI 48824, USA}
\affiliation{Dept. of Physics, University of Alberta, Edmonton, Alberta, Canada T6G 2E1}
\affiliation{Erlangen Centre for Astroparticle Physics, Friedrich-Alexander-Universit{\"a}t Erlangen-N{\"u}rnberg, D-91058 Erlangen, Germany}
\affiliation{Physik-department, Technische Universit{\"a}t M{\"u}nchen, D-85748 Garching, Germany}
\affiliation{D{\'e}partement de physique nucl{\'e}aire et corpusculaire, Universit{\'e} de Gen{\`e}ve, CH-1211 Gen{\`e}ve, Switzerland}
\affiliation{Dept. of Physics and Astronomy, University of Gent, B-9000 Gent, Belgium}
\affiliation{Dept. of Physics and Astronomy, University of California, Irvine, CA 92697, USA}
\affiliation{Karlsruhe Institute of Technology, Institute for Astroparticle Physics, D-76021 Karlsruhe, Germany }
\affiliation{Karlsruhe Institute of Technology, Institute of Experimental Particle Physics, D-76021 Karlsruhe, Germany }
\affiliation{Dept. of Physics, Engineering Physics, and Astronomy, Queen's University, Kingston, ON K7L 3N6, Canada}
\affiliation{Dept. of Physics and Astronomy, University of Kansas, Lawrence, KS 66045, USA}
\affiliation{Department of Physics and Astronomy, UCLA, Los Angeles, CA 90095, USA}
\affiliation{Centre for Cosmology, Particle Physics and Phenomenology - CP3, Universit{\'e} catholique de Louvain, Louvain-la-Neuve, Belgium}
\affiliation{Department of Physics, Mercer University, Macon, GA 31207-0001, USA}
\affiliation{Dept. of Astronomy, University of Wisconsin{\textendash}Madison, Madison, WI 53706, USA}
\affiliation{Dept. of Physics and Wisconsin IceCube Particle Astrophysics Center, University of Wisconsin{\textendash}Madison, Madison, WI 53706, USA}
\affiliation{Institute of Physics, University of Mainz, Staudinger Weg 7, D-55099 Mainz, Germany}
\affiliation{Department of Physics, Marquette University, Milwaukee, WI, 53201, USA}
\affiliation{Institut f{\"u}r Kernphysik, Westf{\"a}lische Wilhelms-Universit{\"a}t M{\"u}nster, D-48149 M{\"u}nster, Germany}
\affiliation{Bartol Research Institute and Dept. of Physics and Astronomy, University of Delaware, Newark, DE 19716, USA}
\affiliation{Dept. of Physics, Yale University, New Haven, CT 06520, USA}
\affiliation{Dept. of Physics, University of Oxford, Parks Road, Oxford OX1 3PU, UK}
\affiliation{Dept. of Physics, Drexel University, 3141 Chestnut Street, Philadelphia, PA 19104, USA}
\affiliation{Physics Department, South Dakota School of Mines and Technology, Rapid City, SD 57701, USA}
\affiliation{Dept. of Physics, University of Wisconsin, River Falls, WI 54022, USA}
\affiliation{Dept. of Physics and Astronomy, University of Rochester, Rochester, NY 14627, USA}
\affiliation{Department of Physics and Astronomy, University of Utah, Salt Lake City, UT 84112, USA}
\affiliation{Oskar Klein Centre and Dept. of Physics, Stockholm University, SE-10691 Stockholm, Sweden}
\affiliation{Dept. of Physics and Astronomy, Stony Brook University, Stony Brook, NY 11794-3800, USA}
\affiliation{Dept. of Physics, Sungkyunkwan University, Suwon 16419, Korea}
\affiliation{Institute of Basic Science, Sungkyunkwan University, Suwon 16419, Korea}
\affiliation{Institute of Physics, Academia Sinica, Taipei, 11529, Taiwan}
\affiliation{Dept. of Physics and Astronomy, University of Alabama, Tuscaloosa, AL 35487, USA}
\affiliation{Dept. of Astronomy and Astrophysics, Pennsylvania State University, University Park, PA 16802, USA}
\affiliation{Dept. of Physics, Pennsylvania State University, University Park, PA 16802, USA}
\affiliation{Dept. of Physics and Astronomy, Uppsala University, Box 516, S-75120 Uppsala, Sweden}
\affiliation{Dept. of Physics, University of Wuppertal, D-42119 Wuppertal, Germany}
\affiliation{DESY, D-15738 Zeuthen, Germany}

\author{R. Abbasi}
\affiliation{Department of Physics, Loyola University Chicago, Chicago, IL 60660, USA}
\author{M. Ackermann}
\affiliation{DESY, D-15738 Zeuthen, Germany}
\author{J. Adams}
\affiliation{Dept. of Physics and Astronomy, University of Canterbury, Private Bag 4800, Christchurch, New Zealand}
\author{J. A. Aguilar}
\affiliation{Universit{\'e} Libre de Bruxelles, Science Faculty CP230, B-1050 Brussels, Belgium}
\author{M. Ahlers}
\affiliation{Niels Bohr Institute, University of Copenhagen, DK-2100 Copenhagen, Denmark}
\author{M. Ahrens}
\affiliation{Oskar Klein Centre and Dept. of Physics, Stockholm University, SE-10691 Stockholm, Sweden}
\author{J.M. Alameddine}
\affiliation{Dept. of Physics, TU Dortmund University, D-44221 Dortmund, Germany}
\author{A. A. Alves Jr.}
\affiliation{Karlsruhe Institute of Technology, Institute for Astroparticle Physics, D-76021 Karlsruhe, Germany }
\author{N. M. Amin}
\affiliation{Bartol Research Institute and Dept. of Physics and Astronomy, University of Delaware, Newark, DE 19716, USA}
\author{K. Andeen}
\affiliation{Department of Physics, Marquette University, Milwaukee, WI, 53201, USA}
% ...

\date{\today}

\begin{abstract}
We present a search for an unstable sterile neutrino by looking for a matter-induced signal in eight years of atmospheric $\nu_\mu$ data collected from 2011 to 2019 at the IceCube Neutrino Observatory. 
Both the (stable) three-neutrino and the 3+1 sterile neutrino models are disfavored relative to the unstable sterile neutrino model, though with $p$-values of 2.5\% and 0.81\%, respectively, we do not observe evidence for 3+1 neutrinos with neutrino decay.  The best-fit parameters for the sterile neutrino with decay model from this study are $\Delta m_{41}^2=6.7^{+3.9}_{-2.5}\,\rm{eV}^2$, $\sin^2 2\theta_{24}=0.33^{+0.20}_{-0.17}$, and $g^2=2.5\pi\pm1.5\pi$, where $g$ is the decay-mediating coupling. The preferred regions from short-baseline oscillation searches are excluded at 90\% C.L.  
\end{abstract}

\collaboration{IceCube Collaboration}
\noaffiliation

\maketitle
\email{analysis@icecube.wisc.edu}

Longstanding anomalies in short-baseline (SBL) neutrino experiments \cite{Athanassopoulos:1997pv,MiniBooNE:2020pnu} have been interpreted in the standard oscillation framework of three known flavors and one or more hypothetical sterile neutrinos, referred to as ``3+N'' models. The ``3+1'' model, which involves only one sterile neutrino, has been extensively studied through global fits to data sets sensitive to vacuum oscillations involving a dominant mass splitting of $\sim 1$ eV$^2$ \cite{Dentler:2018sju, Diaz:2019fwt,Gariazzo:2017fdh}.  These fits find a strong preference for 3+1 over the three neutrino hypothesis \cite{Diaz:2019fwt}. However, the allowed regions from these fits suffer from internal inconsistencies between datasets, referred to as ``tension'' \cite{Maltoni:2003cu, Diaz:2019fwt}.  In particular, no experiment has found evidence of $\nu_\mu$ disappearance, which is expected in a 3+1 model. This is one motivation to reconsider the 3+1 model; another is to evade cosmological bounds on light sterile neutrinos~\cite{Planck:2018vyg,Dolgov:2002wy,Chu:2018gxk,Abazajian:2017tcc,Hagstotz:2020ukm} and possibly resolve the Hubble tension~\cite{Verde:2019ivm,Riess:2019qba,Kreisch:2019yzn,DiValentino:2020vnx,Archidiacono:2016kkh}.

Other explanations for the observed anomalies include misestimation of Standard Model backgrounds in the experiments with anomalies \cite{MiniBooNE:2020pnu,Ericson:2016yjn,Brdar:2021cgb,Giunti:2019sag,Alvarez-Ruso:2015kua}, alternative models that do not involve light sterile neutrinos \cite{Abdallah:2020vgg,Abdullahi:2020nyr,Magill:2018jla,Jones:2019tow,Bertuzzo:2018itn,Fischer:2019fbw,Dentler:2019dhz,deGouvea:2019qre,Gninenko:2009ks,Carena:2017qhd,Masip:2012ke,Dib:2011jh}, and extensions to the 3+1 model that address the internal tension \cite{Bai:2015ztj,Moss:2017pur,Vergani:2021tgc}. In the latter case, models wherein the sterile neutrino is unstable (``3+1+decay'') reduce the tension compared to the 3+1 model \cite{Diaz:2019fwt, Moulai:2019gpi, Vergani:2021tgc}.  However, to be seen as a well-motivated improvement, the 3+1+decay model should be tested through entirely different processes than vacuum oscillations.  

The IceCube Neutrino Observatory has the unique capability of performing such a test. IceCube is a cubic-kilometer neutrino detector buried $1.5$\,km$-2.5$\,km beneath the surface of the Antarctic glacier at the South Pole~\cite{Aartsen:2016nxy}. Muon tracks from charged current (CC) muon \mbox{(anti)neutrino} interactions are reconstructed based on observation of emitted Cherenkov light that is collected by ``Digital Optical Modules'' (DOMs)~\cite{Abbasi:2008aa} arranged in vertical strings on a hexagonal lattice. Specifically, the track-fitting \cite{Aartsen:2013vja} utilizes signals from two detector arrays: 1) the main array of 78 strings spaced 125\,m apart, each carrying 60 DOMs with a vertical separation of 17\,m between them; and 2) the DeepCore~\cite{Collaboration:2011ym} eight-string array, with lateral spacing varying from 42\,m to 72\,m, and vertical DOM separation of 7\,m. 

The existence of an eV-scale sterile neutrino can manifest itself as a resonant, matter-enhanced flavor transition for either muon antineutrinos or muon neutrinos traversing the core of the Earth~\cite{Akhmedov:1988kd,Krastev:1989ix,Chizhov:1998ug,Chizhov:1999az,Akhmedov:1999va,Nunokawa:2003ep}. This causes a deficit of ``up-going'' muon \mbox{(anti)neutrinos} at TeV-scale energies. IceCube cannot distinguish between neutrinos and antineutrinos. Therefore, the only signature is a deficit in the combined muon neutrino and muon antineutrino ($\nu_\mu + \bar{\nu}_\mu$) CC event distribution at TeV energies. A search in the framework of the 3+1 model using eight years of IceCube data has recently been published \cite{meowsprl,meowsprd}. This data set offers an excellent platform to test the hypothesis that the 3+1+decay model provides a better description of the data than the 3+1 model without relying upon the vacuum oscillation signature.

In the 3+1+decay model, the three-neutrino mixing matrix,  ${\rm U}_{\rm{PNMS}}$, which is parametrized by three mixing angles and one CP-violating phase, $\delta_{\rm{CP}}$, is extended by one row and column, adding one sterile flavor state, $\nu_s$, and one heavy mass state, $\nu_4$.  This introduces three new mixing angles, $\theta_{14}$, $\theta_{24}$, and $\theta_{34}$, two new CP-violating phases, $\delta_{14}$ and $\delta_{24}$, and one additional mass splitting, $\Delta m^2_{41}$. Lastly, instability of the fourth mass state is introduced as in Ref.~\cite{Diaz:2019fwt}, governed by the strength of a coupling constant $g$. For non-zero values of $g$, $\nu_4$ can decay into invisible particles beyond the Standard Model, while $g=0$ returns the $3+1$ model. The relationship between this coupling, $g$, the $\nu_4$ mass, $m_4$, and its lifetime, $\tau$, is~\cite{Kim:1990km}
\begin{equation}
\tau = \frac{16\pi}{g^2 m_4}.
\end{equation}
Most of the parameters involved in three-neutrino mixing are well known; these include the light, active neutrino mass splittings and the PMNS matrix elements~\cite{Zyla:2020zbs}. However, this work is insensitive to these parameters as well the neutrino mass ordering and $\delta_{\rm{CP}}$ because, for the relevant neutrino energies ($E_\nu > 100\,$GeV) and baselines (order the diameter of the Earth or smaller), oscillation probabilities between the three active flavors are insignificant. For the present study, the normal mass ordering is assumed and $\delta_{\rm{CP}}$ is assumed to be zero. Furthermore $\theta_{14}$, $\delta_{\rm{14}}$, and $\delta_{\rm{24}}$ are set to zero since they have subleading effects~\cite{Esmaili:2013vza}; $\theta_{34}$ is set to zero as this yields conservative results~\cite{Esmaili:2013vza,Lindner:2015iaa} and $m_1$ is set to zero since only mass differences are relevant. This leaves three free parameters in the model to be tested: $\Delta m^2_{41}$, $\sin^2 2\theta_{24}$, and $g^2$. It is assumed that $\theta_{24} < \pi/4$, which causes the resonance to appear in the antineutrino flux; larger values of $\theta_{24}$ are heavily constrained~\cite{Diaz:2019fwt} and since there are more atmospheric neutrinos than antineutrinos~\cite{Gaisser:2011cc}, this choice is also conservative.

At IceCube, the $\nu_\mu$ and $\bar\nu_\mu$ disappearance probabilities vary as a function of energy and zenith angle ($\theta_{\rm{z}}$), where $\cos\theta_{\rm{z}}=0$ corresponds to neutrinos arriving from the horizon and $\cos\theta_{\rm{z}}=-1$ corresponds to neutrinos arriving from the direction of the North Pole. This study uses a dataset collected over a livetime of 2786 days and an event selection that has been described in detail in Refs.~\cite{meowsprd, Axani:2019sbk}. The predicted resonance occurs at TeV scales, hence the analysis focuses on muons from  CC neutrino interactions with energies between 500\,GeV and 10\,TeV.  Relevant neutrino interactions occur below or within IceCube. Because the signature of the analysis relies on matter effects in the Earth, this analysis requires muon tracks to have up-going zenith angle ($-1.0<\cos\theta_{\rm{z}}<0.0$). The angular resolution of the tracks, $\sigma_{\cos \theta_{\rm{z}}}$, lies between 0.005 and 0.015, and the track energy resolution, $\sigma_{\log_{10}  E_\mu}$, is $\sim 0.5$~\cite{Aartsen:2013vja}. The selected sample comprises 305,735 CC $\nu_\mu$  and $\bar\nu_\mu$ events.

The expected neutrino flux is primarily atmospheric neutrinos, with approximately a 3\% overall contribution from astrophysical neutrinos, determined by extrapolating from measurements at higher energies ~\cite{Aartsen:2013jdh,Aartsen:2014gkd,Aartsen:2015rwa,Aartsen:2016xlq,Aartsen:2018vez,IceCube:2020wum,IceCube:2020acn}. The atmospheric flux arises predominantly from the decays of kaons and pions, and to a much lesser extent, muons, in cosmic-ray air showers~\cite{Fedynitch:2015zma}. The decays of heavier mesons contribute minimally to the atmospheric flux in the energy range relevant to this analysis~\cite{Garzelli:2015psa,Gauld:2015kvh,Gauld:2015yia,Bhattacharya:2016jce,Garzelli:2016xmx,Goncalves:2006ch,Enberg:2008te}. The atmospheric and astrophysical fluxes fall steeply with energy, with spectral indices of approximately $-3.7$ and $-2.5$, respectively~\cite{IceCube:2010jrf, IceCube:2020acn}. 

The physics under study affects the flavors of the neutrinos as they propagate through the Earth.  This is described using the  \texttt{nuSQuIDS} neutrino evolution code~\cite{Delgado:2014kpa,Arguelles:2021twb} which 
accounts for both coherent and non-coherent interactions~\cite{GonzalezGarcia:2005xw,Formaggio:2012cpf,Gandhi:1998ri,Zhou:2019frk,Zhou:2019vxt,Garcia:2020jwr}, as well as tau neutrino regeneration~\cite{Halzen:1998be, Dutta:2000hh}.  This analysis uses \texttt{nuSQUIDSDecay}, which incorporates the effect of $\nu_4$ decay~\cite{Moss:2017pur}.  The Earth density profile is parameterized by the spherically-symmetric PREM model~\cite{Dziewonski:1981xy}. The CSMS~\cite{CooperSarkar:2011pa} neutrino-nucleon cross section is used to describe the CC interactions below and within the detector.

This analysis builds on the 3+1 analysis in Ref.~\cite{meowsprd}. The data are binned in reconstructed muon energy and $\cos \theta_z$, and a modified Poisson likelihood that accounts for finite simulation statistics is used to evaluate the data given sterile neutrino parameters~\cite{Arguelles:2019izp}. Eighteen systematic effects related to the atmospheric and astrophysical flux, detector, and cross section uncertainties are incorporated into the likelihood function as nuisance parameters; these are described further in Ref.~\cite{meowsprd}. The treatment of most systematic uncertainties is unchanged. The dominant category of uncertainties had been identified as those associated with the atmospheric neutrino flux.

Two improvements were made over the 3+1 analysis: 1) the uncertainty in the atmospheric neutrino flux corresponding to the uncertainty in the production of charged mesons in atmospheric showers is calculated using atmospheric data from the NASA Atmospheric InfraRed Sounder satellite~\cite{airs}, rather than the atmospheric model from Ref.~\cite{atmosphere1976national}; and 2) the astrophysical and prompt neutrino fluxes are calculated using a corrected depth setting of the glacial ice, compared to Ref.~\cite{meowsprd}, which had little impact on the current or previous results. Combined, these changes increase the likelihood of the data for the three-neutrino model and best-fit $3+1$ model by, respectively, 0.09 and 0.18 log-likelihood units.

Both a frequentist parameter estimation and a point-wise Bayesian model comparison~\cite{Gariazzo:2018meg} are performed, following the same procedure as in Ref.~\cite{meowsprd}. The likelihood function and Bayes factor~\cite{RevModPhys.83.943} are evaluated over a grid scan of the three physics parameters -- $\Delta m_{41}^2$, $\sin^2 2\theta_{24}$, and $g^2$ -- where $\Delta m_{41}^2$ and $\sin^2 2\theta_{24}$ are sampled $\log$-uniformly with ten samples per decade in the ranges $0.01 - 47\,\rm{eV}^2$ and $0.01 - 1.0$, respectively, and the parameter $g^2$ is sampled in steps of $\frac{\pi}{2}$ in the range $0-4\pi$.

\begin{figure}[th!]
\centering
\includegraphics[width=\columnwidth]{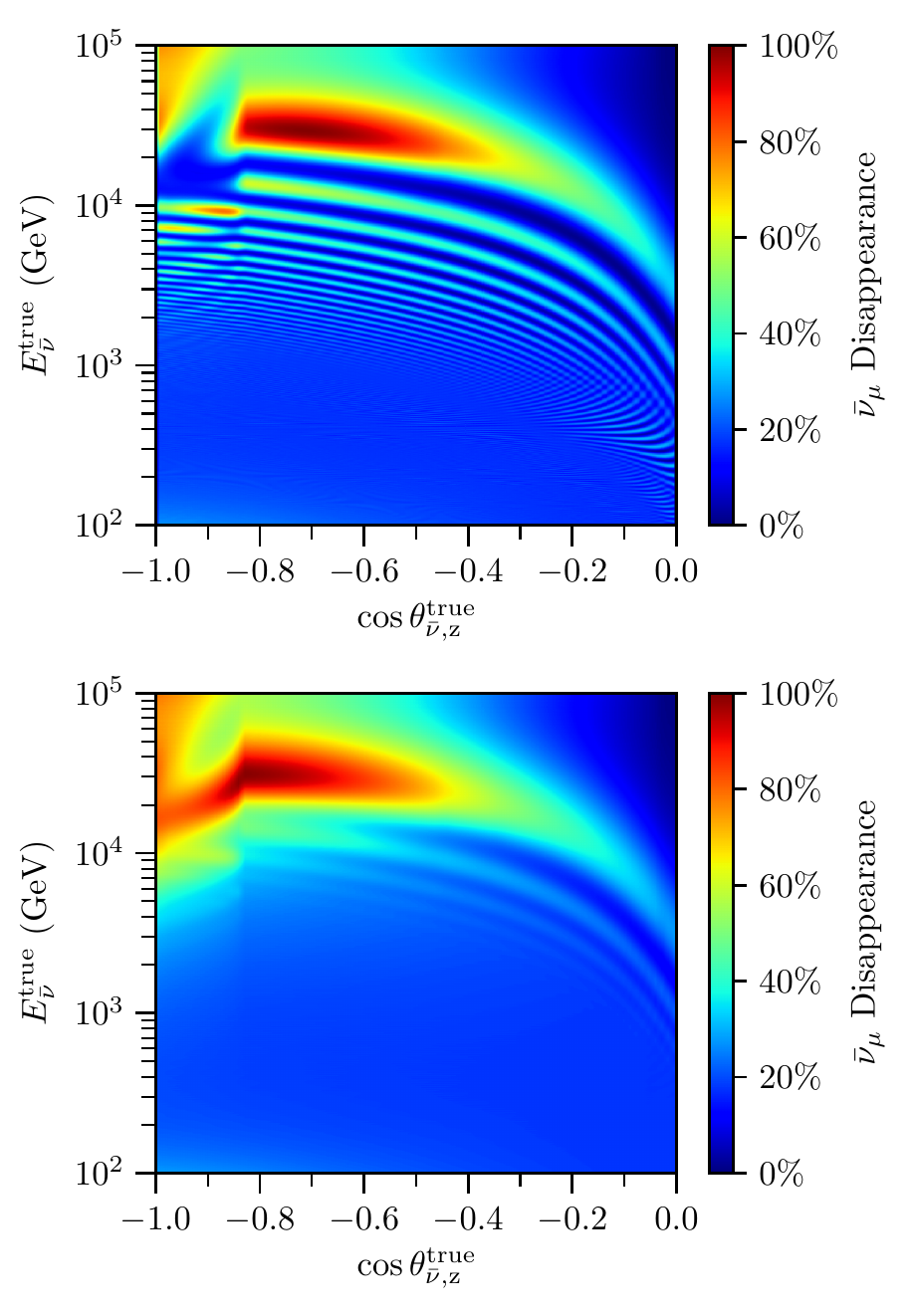} 
\caption{Muon antineutrino disappearance probabilities for the sterile parameters $\Delta m_{41}^2 = 6.7$ eV$^2$, $\sin^2 2 \theta_{24} = 0.33 $ and two values of $g^2$. Top: $g^2 = 0$, which corresponds to infinite $\nu_4$ lifetime, \textit{i.e.} the $3+1$ model. Bottom: $g^2 = 2.5 \pi$; this is the best-fit point.}
\label{fig::oscillograms}
\end{figure}

The best-fit parameters are found to be $\Delta m_{41}^2=6.7^{+3.9}_{-2.5}\,\rm{eV}^2$, $\sin^2 2\theta_{24}=0.33^{+0.20}_{-0.17}$, and $g^2=2.5\pi\pm1.5\pi$. The $\bar\nu_\mu$ disappearance probabilities are given in  Fig.~\ref{fig::oscillograms} for the parameters $\Delta m_{41}^2 = 6.7\,\rm{eV}^2$ and $\sin^2 2 \theta_{24} = 0.33$, and for two values of $g^2$; the top panel shows the situation for $g^2 = 0$, which corresponds to the $3+1$ model, while the bottom panel is for the case $g^2 = 2.5\pi$. The bottom panel represents the best-fit point of the frequentist analysis. Muon neutrino disappearance probabilities do not feature the resonant deficit and make sub-leading contributions to the sterile signature, so they are not shown.

\begin{figure}[t]
\centering
\includegraphics[width=\columnwidth]{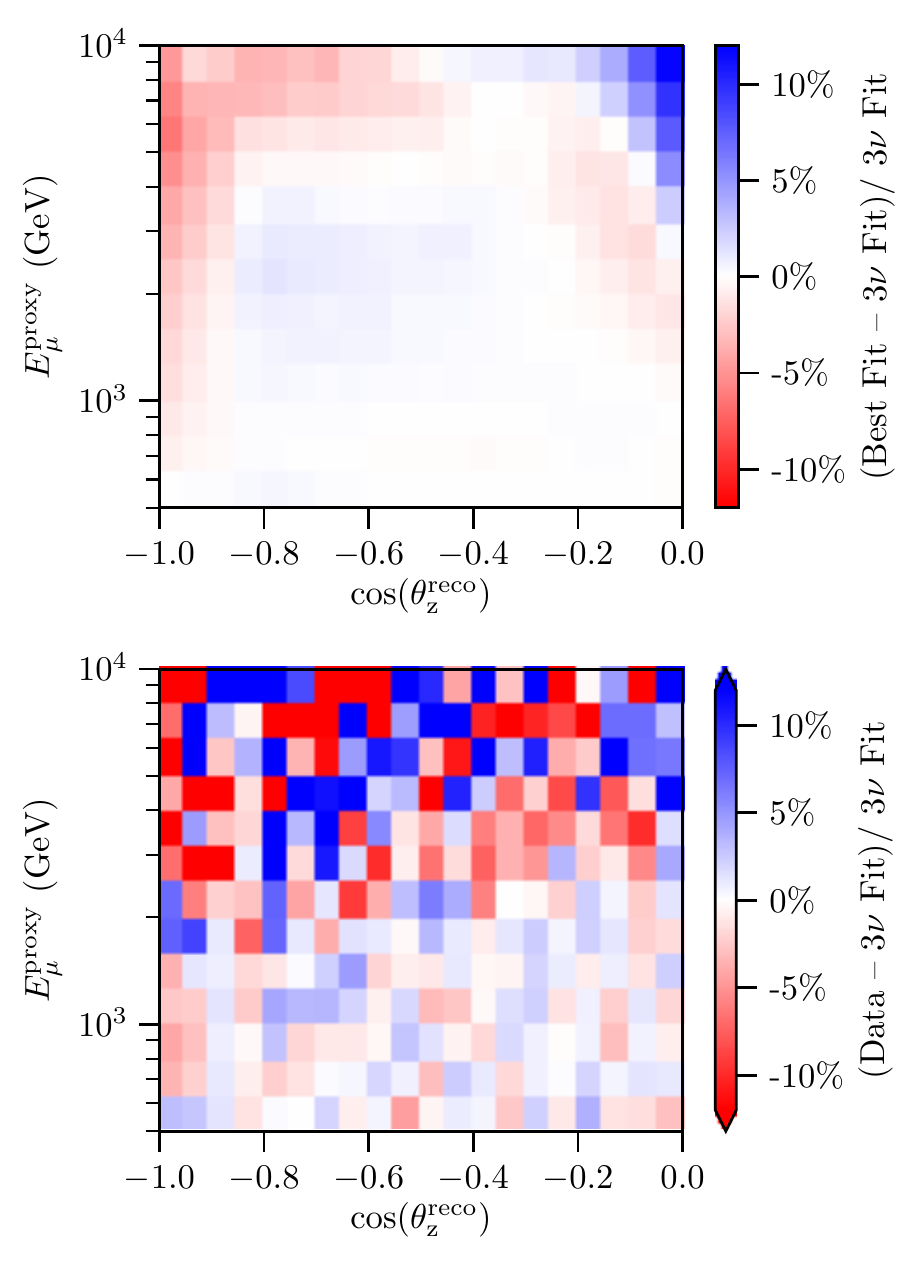}
\caption{Top: Comparison of best-fit signal expectation to the three-neutrino fit. Bottom: Comparison of the binned data to the three-neutrino fit. Distributions are binned in reconstructed muon energy ($E_\mu^{\rm{Proxy}}$) and cosine zenith angle.}
\label{fig::true_signal}
\end{figure}

The best-fit signal expectation and data are both compared to the three-neutrino model expectation in Fig.~\ref{fig::true_signal}. In these plots, both the signal expectation and the three-neutrino model expectation include systematic uncertainties estimated adopting the respective physics parameters.  Both the data and the best-fit signal shapes have a deficit of events for through-going neutrinos at the highest energies and a relative excess for horizon-skimming events at the highest energies. The fit values of all systematic uncertainties are within 1$\sigma$ of their prior centers, with the exception of the cosmic-ray spectral index. The fit value of this systematic uncertainty deviates by 2.4$\sigma$, which is similar to both the result from the 3+1 search~\cite{meowsprl,meowsprd}, as well as the fit value assuming no sterile neutrino.   

\begin{figure}[t!]
\centering
\includegraphics[width=\columnwidth]{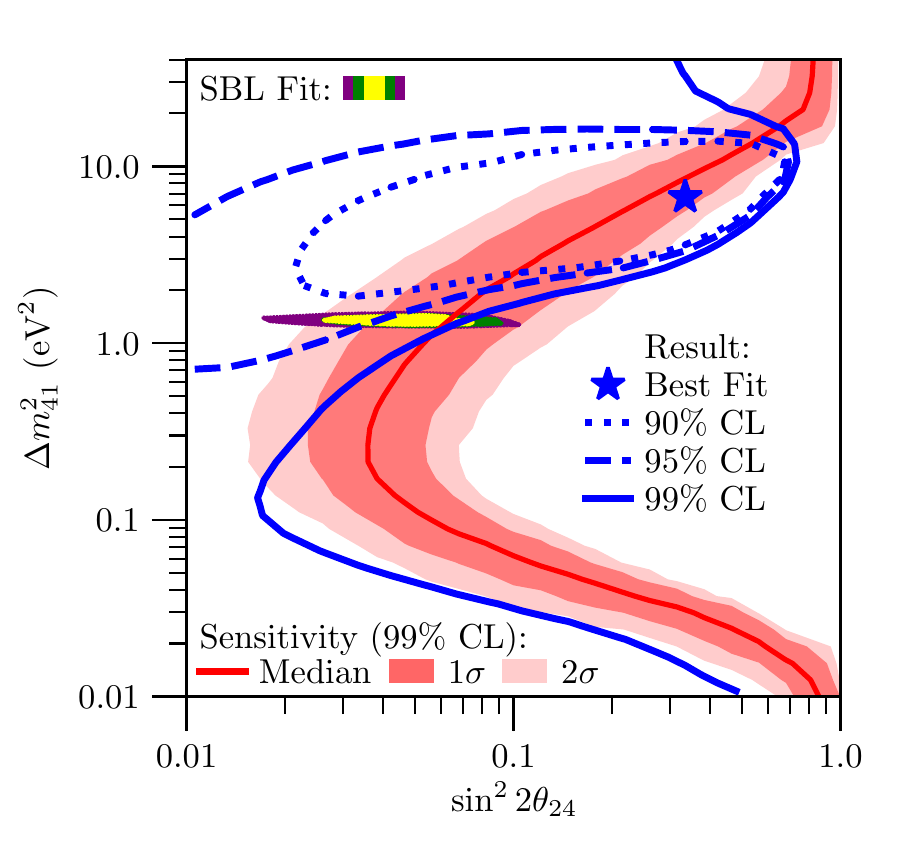}
\caption{The result of the frequentist analysis for $g^2 = 2.5\pi$. The 90\%, 95\%, and 99\% C.L. contours are shown as blue dotted, dashed, and solid curves, respectively. The best-fit point is marked with a blue star. The median sensitivity at 99\% C.L., determined from 300 simulated datasets, is shown as a red curve. The medium and light pink bands indicate the $1\sigma$ and $2\sigma$ regions for the sensitivity. The 2D projection of the SBL fit results from~\cite{Diaz:2019fwt} for the range $2.25\pi \leq g^2 \leq 2.75\pi$ at 90\% C.L., 95\% C.L., and 99\% C.L. are shown as the solid yellow, green, and purple islands around $\Delta m^2_{41} = 1.3\,\rm{eV}^2$.}
\label{fig::frequentist}
\end{figure}

The frequentist confidence regions sliced at the best-fit value of $g^2$ are shown in Fig.~\ref{fig::frequentist}. The contours are drawn assuming Wilks' theorem and three degrees of freedom (DOF). Fits to simulated datasets for several points in the parameter space showed the effective DOF was consistent with three or fewer. The slices of the confidence regions for the other values of $g^2$ are approximately the same in the 2D space of [$\Delta m_{41}^2$, $\sin^2 2\theta_{24}$], with two deviations: the 90\% C.L. region for $g^2 = 0$ excludes any point with $\sin^2 2\theta_{24} \gtrsim 0.2$, and above $\Delta m_{41}^2 \sim 7\,\rm{eV}^2$, the confidence regions extend to higher $\Delta m_{41}^2$ values for larger values of $g^2$. This is shown in the Supplementary Material. The null hypothesis of only three neutrinos is disfavored in favor of this $3+1+\textrm{decay}$ model with a $p$-value of 2.5\%. This $p$-value was obtained using Wilks' theorem and 2.86 effective DOF. The effective DOF was determined by fitting 300 simulated datasets generated assuming the null hypothesis.

\begin{figure}[!t]
\centering
\includegraphics[width=\columnwidth]{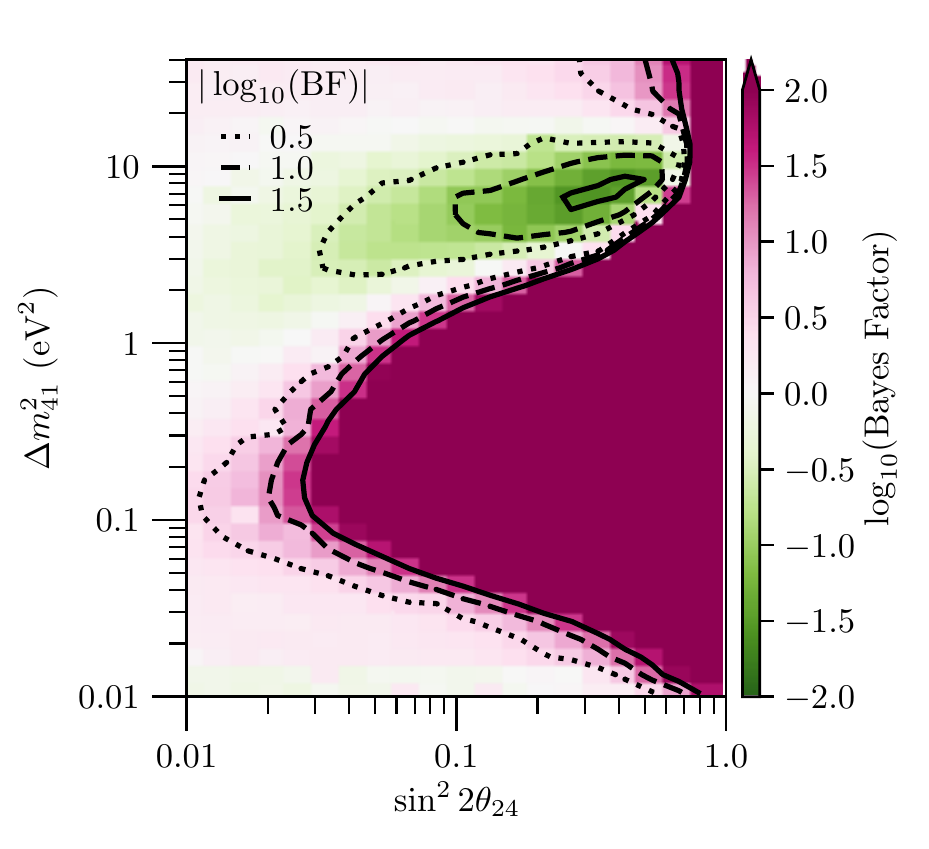}

\caption{The result of the Bayesian analysis for $g^2 = 2.5\pi$. The color indicates the logarithm of the Bayes factor with respect to the three-neutrino model; magenta regions have strong preference for the three-neutrino model, while green regions have preference for the sterile neutrino model. The dotted, dashed, and solid black contours correspond to $\log_{10}$(BF) equaling $\pm$0.5, $\pm$1.0, and $\pm$1.5, respectively.}
\label{fig::bayes}
\end{figure}

The Bayesian analysis finds the best model to have the parameters $\Delta m_{41}^2 = 6.7\:\rm{eV}^2$, $\sin^2 \theta_{24} = 0.33$, and $g^2 = 1.5\pi$; this model has a Bayes factor (BF) with respect to the three-neutrino model of 0.025. The Bayes factor of the frequentist best-fit point is 0.027. The Bayesian result for $g^2 = 2.5\pi$ is shown in Fig.~\ref{fig::bayes}. As with the frequentist confidence regions, Bayesian preferred regions sliced at the varying values of $g^2$ are very similar, with a few exceptions. For $g^2 = 0$, the region $\log_{10}(\rm{BF}) \leq - 0.5$ excludes points with $\sin^2 2\theta_{24} \gtrsim 0.2$. The regions with $\log_{10}(\rm{BF}) = - 1.5$ only occur for $1.5 \pi \leq g^2 \leq 2.5 \pi$. This is shown in the Supplementary Material.

\begin{figure}[!t]
\centering
\includegraphics[width=\columnwidth]{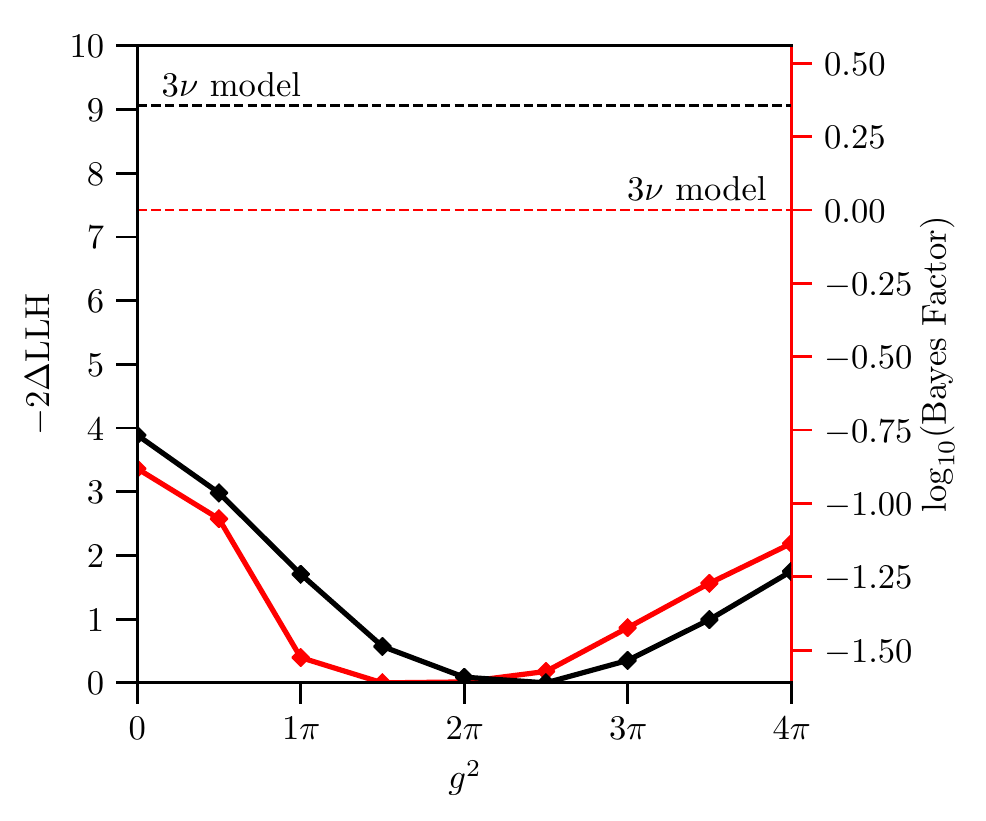}
\caption{The frequentist and Bayesian results profiled over two of the sterile parameters, $\Delta m_{41}^2$ and $\sin^2 2\theta_{24}$. The frequentist test statistic, $-2 \Delta \rm{LLH}$ is shown in black and is plotted on the left $y$-axis. The logarithm of the Bayes factor is shown in red and is plotted on the right $y$-axis. The diamond markers joined by the thick lines show the results for the sterile model as a function of the third sterile parameter, $g^2$. The results for the null hypothesis -- that there are only three neutrino species -- are shown in the dashed horizontal lines at the top of the plot.}
\label{fig::g2}
\end{figure}

The frequentist and Bayesian results profiled over the parameters $\Delta m_{41}^2$ and $\sin^2 2\theta_{24}$ are shown in Fig.~\ref{fig::g2}. Both analyses find some preference for non-zero $g^2$. In the frequentist analysis, $g^2 = 0$ is disfavored in favor of non-zero $g^2$ with a $p$-value of 0.81\%. This $p$-value was obtained using Wilks' theorem and 0.26 effective DOF. The effective DOF was determined by fitting 500 simulated datasets generated assuming the best-fit $3+1$ parameters, \textit{i.e.}, fixing $g^2 = 0$.

The 95\% C.L. allowed region found in this work overlaps that of the SBL fits, as seen in Fig.~\ref{fig::frequentist}. This overlap occurs to some extent for all non-zero values of $g^2$, but is larger for $g^2$ values above $\pi$. This overlap remains fixed in $\Delta m_{41}^2$ and $\sin^2 2\theta_{24}$ for varying $g^2$. At and above $g^2 = \pi$, there is some overlap between the 95\% C.L. region of this work and the 90\% C.L. allowed region from the SBL fits.

In conclusion, we have found no substantive evidence for the 3+1+decay model. The null hypothesis of only three neutrinos is disfavored with a $p$-value of 2.5\%, and the 3+1 model disfavored with a $p$-value of 0.81\%. The best-fit parameters are $\Delta m_{41}^2=6.7^{+3.9}_{-2.5}\,\rm{eV}^2$, $\sin^2 2\theta_{24}=0.33^{+0.20}_{-0.17}$, and $g^2=2.5\pi\pm1.5\pi$. While we have reported valuable new input to global studies, further work is needed to clarify the picture.

\section{Acknowledgements\label{sec:ack}}

\begin{acknowledgments}
The IceCube collaboration acknowledges the significant contributions to this manuscript from the Massachusetts Institute of Technology, University of Texas at Arlington, and Harvard University groups.
The authors gratefully acknowledge the support from the following agencies and institutions: 
USA {\textendash} U.S. National Science Foundation-Office of Polar Programs,
U.S. National Science Foundation-Physics Division,
U.S. National Science Foundation-EPSCoR,
Wisconsin Alumni Research Foundation,
Center for High Throughput Computing (CHTC) at the University of Wisconsin{\textendash}Madison,
Open Science Grid (OSG),
Extreme Science and Engineering Discovery Environment (XSEDE),
Frontera computing project at the Texas Advanced Computing Center,
U.S. Department of Energy-National Energy Research Scientific Computing Center,
Particle astrophysics research computing center at the University of Maryland,
Institute for Cyber-Enabled Research at Michigan State University,
and Astroparticle physics computational facility at Marquette University;
Belgium {\textendash} Funds for Scientific Research (FRS-FNRS and FWO),
FWO Odysseus and Big Science programmes,
and Belgian Federal Science Policy Office (Belspo);
Germany {\textendash} Bundesministerium f{\"u}r Bildung und Forschung (BMBF),
Deutsche Forschungsgemeinschaft (DFG),
Helmholtz Alliance for Astroparticle Physics (HAP),
Initiative and Networking Fund of the Helmholtz Association,
Deutsches Elektronen Synchrotron (DESY),
and High Performance Computing cluster of the RWTH Aachen;
Sweden {\textendash} Swedish Research Council,
Swedish Polar Research Secretariat,
Swedish National Infrastructure for Computing (SNIC),
and Knut and Alice Wallenberg Foundation;
Australia {\textendash} Australian Research Council;
Canada {\textendash} Natural Sciences and Engineering Research Council of Canada,
Calcul Qu{\'e}bec, Compute Ontario, Canada Foundation for Innovation, WestGrid, and Compute Canada;
Denmark {\textendash} Villum Fonden and Carlsberg Foundation;
New Zealand {\textendash} Marsden Fund;
Japan {\textendash} Japan Society for Promotion of Science (JSPS)
and Institute for Global Prominent Research (IGPR) of Chiba University;
Korea {\textendash} National Research Foundation of Korea (NRF);
Switzerland {\textendash} Swiss National Science Foundation (SNSF);
United Kingdom {\textendash} Department of Physics, University of Oxford.
\end{acknowledgments}

\bibliography{main}

\pagebreak
\clearpage
\onecolumngrid
\appendix
\ifx \standalonesupplemental\undefined
\setcounter{page}{1}
\setcounter{figure}{0}
\setcounter{table}{0}
\setcounter{equation}{0}
\fi

\renewcommand{\thepage}{Supplemental Material-- S\arabic{page}}
\renewcommand{\figurename}{SUPPL. FIG.}
\renewcommand{\tablename}{SUPPL. TABLE}
\renewcommand{\theequation}{A\arabic{equation}}
\clearpage
\begin{center}
\textbf{\large Supplemental Material}
\end{center}

\section{\label{sec:supplemental}Supplementary Material: Frequentist and Bayesian complete scans}
\begin{figure*}[h!]
\centering
\includegraphics[width=0.75\textwidth]{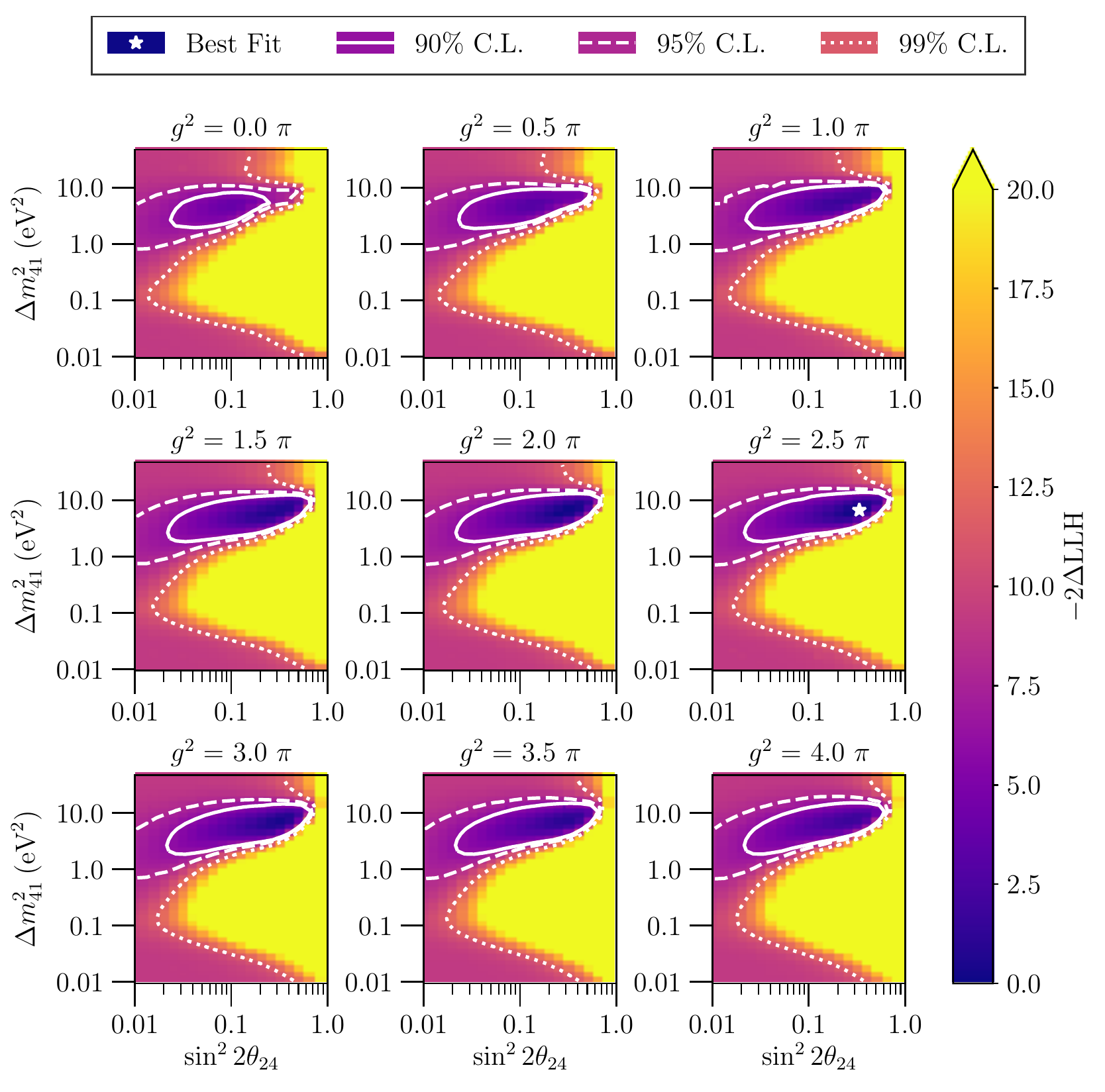}
\caption{Frequentist results for all scanned values of $g^2$. Each of the nine panels corresponds to a fixed value of $g^2$. In each panel, $-2\Delta\textrm{LLH}$ is plotted as a function of $\Delta m^2_{41}$ and $\sin^2 2\theta_{24}$. The solid, dashed, and dotted contours white correspond to 90\%, 95\%, and 99\% C.L. The best-fit point is marked with a white star. Systematic uncertainties are fit for each point ($\Delta m_{41}^2,\; \sin^2 2 \theta_{24}, \: g^2$) in the scan.}
\label{fig::freq_9panels}
\end{figure*}

\begin{figure*}
\centering
\includegraphics[width=0.75\textwidth]{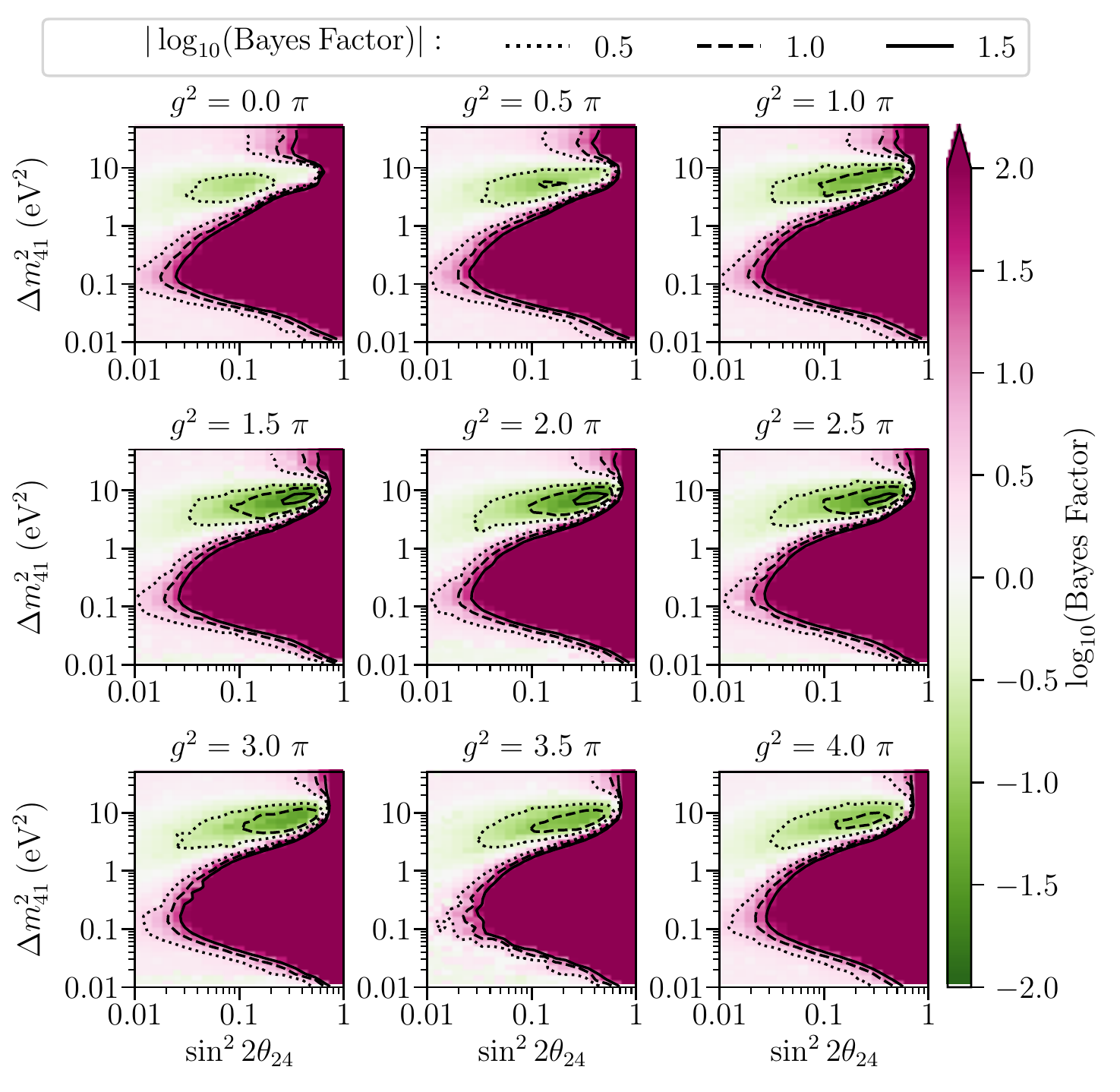}
\caption{Bayesian results for all scanned values of $g^2$. Each of the nine panels corresponds to a fixed value of $g^2$. In each panel, $\log_{10}(\textrm{BF})$, where $\textrm{BF}$ is the Bayes factor with respect to the three-neutrino model, is plotted as a function of $\Delta m^2_{41}$ and $\sin^2 2\theta_{24}$. The dotted, dashed, and solid black contours correspond to $\log_{10}(\textrm{BF})$ values of $\pm 0.5$, $\pm 1.0$, and $\pm 1.5$. Green areas are where the sterile model is preferred; magenta areas are where the null (three-neutrino) hypothesis is preferred. Systematic uncertainties are marginalized for each point ($\Delta m_{41}^2,\; \sin^2 2 \theta_{24}, \: g^2$) in the scan.} 
\label{fig::bayes_9panels}
\end{figure*}

This section contains the complete scan results, that is, the results for all nine scanned values of $g^2$, for both the frequentist and Bayesian analyses. The frequentist results are given in Suppl. Fig.~\ref{fig::freq_9panels} and the Bayesian results are given in Suppl. Fig.~\ref{fig::bayes_9panels}.

\end{document}